\documentclass[aps,pra,twocolumn,showpacs,showkeys,groupedaddress]{revtex4}

\usepackage{graphicx}
\usepackage{amsthm,amssymb,amsmath}
\usepackage[T1]{fontenc}

\begin{document}


\title{High-visibilty two-photon interference at a telecom wavelength\\ using picosecond regime separated sources}

\author{Pierre Aboussouan}
\affiliation{Laboratoire de Physique de la Matière Condensée, CNRS UMR 6622, Université de Nice -- Sophia Antipolis, Parc Valrose, 06108 Nice Cedex 2, France.
}
\author{Olivier Alibart}%
\affiliation{Laboratoire de Physique de la Matière Condensée, CNRS UMR 6622, Université de Nice -- Sophia Antipolis, Parc Valrose, 06108 Nice Cedex 2, France.
}
\author{Daniel B. Ostrowsky}%
\affiliation{Laboratoire de Physique de la Matière Condensée, CNRS UMR 6622, Université de Nice -- Sophia Antipolis, Parc Valrose, 06108 Nice Cedex 2, France.
}
\author{Pascal Baldi}%
\affiliation{Laboratoire de Physique de la Matière Condensée, CNRS UMR 6622, Université de Nice -- Sophia Antipolis, Parc Valrose, 06108 Nice Cedex 2, France.
}
\author{Sébastien Tanzilli}%
\email{sebastien.tanzilli@unice.fr}
\affiliation{Laboratoire de Physique de la Matière Condensée, CNRS UMR 6622, Université de Nice -- Sophia Antipolis, Parc Valrose, 06108 Nice Cedex 2, France.
}

\date{\today}

\begin{abstract}
We report on a two-photon interference experiment in a quantum relay configuration using two picosecond regime PPLN waveguide based sources emitting paired photons at 1550\,nm. The results show that the picosecond regime associated with a guided-wave scheme should have important repercussions for quantum relay implementations in real conditions, essential for improving both the working distance and the efficiency of quantum cryptography and networking systems. In contrast to already reported regimes, namely femtosecond and CW, it allows achieving a 99\% net visibility two-photon interference while maintaining a high effective photon pair rate using only standard telecom components and detectors.
\end{abstract}

\maketitle

For the realization of quantum networks, interference between photons produced by independent sources is necessary. Photon coalescence (or two-photon interference) lies at the heart of quantum operations and is seen as a first step towards achieving both teleportation~\cite{Landry_Teleswisscom_07} and entanglement swapping~\cite{DeRiedmatten_swapping_05,Yang_Syncro_Indep_06,Kaltenbaek_Inter_Indep_09,Halder_Ent_Indep_07,Takesue_Ent_swap_09,Jang_inter_CW_indep_09}.
This effect has been extensively studied theoretically~\cite{Rbook,LegeroRempe_SPcharac_coalecence_03} and experimentally, initially based on two photons coming from a single down conversion source and therefore sharing a common past~\cite{HOM1987,Rarity_dip_89,Halder_2times25km_05}. However, experiments involving truly independent photons represent an important challenge for achieving longer quantum links by means of quantum relays~\cite{Collins_QRelays_05}. In this frame, it has been demonstrated theoretically that a two-photon interference net visibility of at least 95\% is required for practical implementations using currently available photon pair sources and multimode quantum memories~\cite{Simon07}. Reaching such a high visibility therefore appears to be a hard task since a perfect synchronization between independent sources is necessary to prevent any kind of distinguishability between the interfering photons. Several papers focusing on the synchronization issue have demonstrated that entanglement swapping with fully independent sources is in principle feasible in the femtosecond~\cite{Yang_Syncro_Indep_06,Kaltenbaek_Inter_Indep_09} and CW~\cite{Halder_Ent_Indep_07,Jang_inter_CW_indep_09} regimes. Unfortunately, beyond the fundamental interest, the reported interference visibilities remain either far from 95\% or show very low overall photon pair rate.
For instance, the best visibility reported so far in the femtosecond regime has been obtained by compensating the synchronization-induced temporal distinguishability by dramatically increasing the photons coherence time up to a few picoseconds at the expense of the overall brightness~\cite{Kaltenbaek_Inter_Indep_09}. Since laser cavities can easily be synchronized to subpicosecond accuracy~\cite{Hudson_synchro_06,Kim_synchro_08}, the study of the picosecond regime~\cite{Takesue_Ent_swap_09,Fulconis_TwoPhotInterf_07}, its associated filter bandwidths, and the type of photon pair generators, becomes of prime interest to ensure simultaneously a high degree of indistinguishability and a high overall brightness.

Here, we demonstrate that to achieve this, the picosecond regime should provide an efficient trade-off enabling near-perfect two-photon interference and high effective photon pair rates, when associated with standard components available from the telecommunications industry. More precisely, we realized an experiment based on a single picosecond pump laser and two periodically poled lithium niobate (PPLN) waveguides emitting paired photons around 1550 nm to approximate independent sources. We report the highest two-photon interference net visibility, i.e. of 99\%, ever demonstrated in a  configuration extendable to quantum relays. Such a proof-of-principle emphasizes why guided-wave technology, in the picosecond regime, should lead to realistic quantum relay schemes, namely by offering an reduced-constraint solution for the synchronization issue when two completely independent lasers are employed. In the following, we briefly introduce the entanglement swapping based quantum relay scheme and discuss why the picosecond regime is a valuable trade-off. We then focus on our experimental demonstration. Finally, we detail a comparative study of performance with similar reported experiments.

\begin{figure}[h]
\scalebox{.48}{\includegraphics{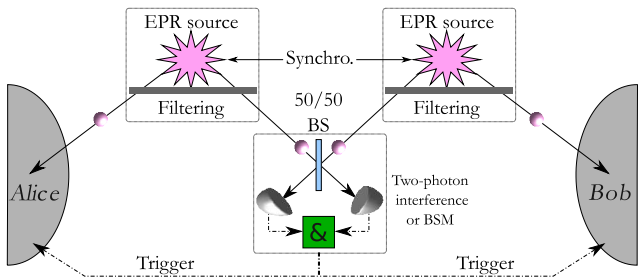}}
\caption{Schematics of a quantum relay involving two pairs of entangled photons emitted by two synchronized sources (EPR). The two inner photons are sent to a 50/50 BS where a Bell state measurement (BSM), based on two-photon interference, is performed. Using a dedicated detection, entanglement can be swapped from these photons to the outer ones making Alice and Bob connected by entanglement, as if they had each received one photon from an entangled pair directly. The BSM serves as a trigger for Alice and Bob's detectors. It therefore enables reducing the SNR of the overall quantum link and further increasing the maximum achievable distance.}
\label{Fig_basics_synchro_filter}
\end{figure}

A basic scheme of a quantum relay based on entanglement swapping is given in~\figurename{\ref{Fig_basics_synchro_filter}} (see caption for details). For long distance quantum communication links, the preferred qubit carriers are photons at 1550\,nm allowing the users, namely Alice and Bob, to take advantage of standard optical fibers for distribution purposes.
Theoretically, the two inner photons can come from any type of source, provided they are identical at the beam-splitter (BS) (pre-selection) or at the detectors (post-selection).
From the experimental side, we have to compare their coherence time, $\tau_{\mbox{\footnotesize coh}}$, to the time uncertainty, $t_{\mbox{\footnotesize uncert}}$, within which they are created (i.e. the pulse duration of the pump laser(s)) or are detected (detector's timing jitters), which can be written as:
\begin{equation}
\tau_{\mbox{\footnotesize coh}}\geq t_{\mbox{\footnotesize uncert}}\label{temp}
\end{equation}
Suitable bandpass filters are therefore employed to achieve optimal interference visibilities. Up to now, this issue has been addressed using different approaches based on pulsed or CW lasers (see also Table~\ref{comparison} for comparison).

On the one hand, the first experiments reported two independent crystals pumped by femtosecond lasers, since this regime allows working with broadband filters, i.e. on the order of a few nm~\cite{DeRiedmatten_swapping_05}. However, the large photon bandwidths makes them more prone to chromatic and polarization dispersions in optical fibers and leads to reported visibilities below 85\%~\cite{DeRiedmatten_Qint_SpatialSeparated_03}. Moreover when two independent lasers are used, the two related laser cavities have to be properly synchronized so as to be identical within the pump pulse duration ($\sim$100\,fs)~\cite{Yang_Syncro_Indep_06,Kaltenbaek_Inter_Indep_09}. This can be achieved using phase-locked loops or atomic clocks and dedicated electronics~\cite{Shelton01,dossierlaser}. Any remaining jitter between the two lasers has to be compensated by narrowing the filters which implies a substantial reduction of the pair production rate for equivalent powers~\cite{Kaltenbaek_Inter_Indep_09}.

On the other hand, the development of narrow-bandwidth fiber Bragg grating (FBG) filters allows using two independent CW lasers stabilized against an atomic transition~\cite{Halder_Ent_Indep_07,Takesue_Ent_swap_09}. Since this regime does not provide any reference clock, the timing function is transferred to the coincidence detection after the BS. To ensure a high quality interference, the photons coherence time has to be longer than the jitter of the detectors. Experimentally, this is made possible using a suitable combination of low-jitter detectors (based on superconducting or up-conversion technologies), on the order of 70\,ps, associated with ultra-narrow FBG filters, on the order of 10\,pm~\cite{Halder_Ent_Indep_07}. In such a situation, one gets rid of the synchronization but the price to pay is rather high in terms of sensitive fluctuations of the filters central wavelengths, low experimental rates due to the narrow filters coupled to the randomness of entangling photons by coincidence detection, and high statistical fluctuations due to low count rates. All these issues prevent reaching high visibilities.

Between these two extremes, the picosecond regime appears to be an efficient compromise. First, the timing condition~(\ref{temp}) is easily met by using FBG filters showing much larger bandwidths than in CW, i.e. on the order of 100\,pm~\cite{Fulconis_TwoPhotInterf_07}. Second, this allows taking advantage of off-the-shelf InGaAs avalanche photodiodes (APD). Last, when compared to the femtosecond regime, narrower bandwidths loosen the constraints on both path-length stabilization for the interfering photons and laser synchronization to subpicosecond accuracy~\cite{Hudson_synchro_06,Kim_synchro_08} to obtain high-visibility Hong-Ou-Mandel interference. All these key points motivate investigating this regime.

\begin{figure}[h]
\scalebox{.6}{\includegraphics{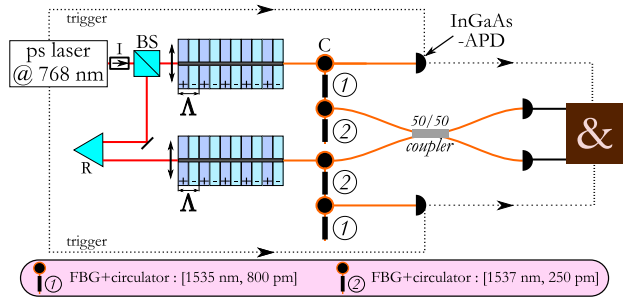}}
\caption{Coalescence experiment based on two PPLN waveguides pumped by a single picosecond laser. A very weak part of this beam is sent to a silicon APD (not represented) employed as a laser clock random divider. Its electrical output is used to trigger four InGaAs-APDs. $\Lambda$: poling period of the PPLN waveguides; I: isolator; R: retro-reflector; C: circulator; \&: AND-gate. The InGaAs-APDs feature 10\% quantum efficiency and a dark count probability of about $10^{-5}$\,ns$^{-1}$.}
\label{Fig_HOM_setup}
\end{figure}

Assuming synchronization is fairly feasible, we therefore focused our efforts at increasing the overall spectral brightness by employing a picosecond pump laser associated with state-of-the-art waveguide sources emitting at telecom wavelengths. As shown in \figurename{\ref{Fig_HOM_setup}}, the pump laser (Coherent MIRA 900-D) provides 1.2\,ps-duration, time-bandwidth limited ($\Delta \lambda_{p}=$ 0.25\,nm) pulses, at the wavelength of $\lambda_{p}=$ 768\,nm and at a repetition rate of 76\,MHz. The pulses are sent to a BS whose outputs are directed towards two 10\,mm-long PPLN waveguides fabricated in our laboratory using the soft-proton exchange technique~\cite{Tanzilli_PPLNW_02}. These devices are single mode at telecom wavelengths enabling only one nonlinear process to occur avoiding any additionnal background noise within the bandwidth of interest which is a key advantage compared to four-wave mixing sources~\cite{Fulconis_TwoPhotInterf_07,Takesue_Ent_swap_09} limited by Raman background photons. Both samples opto-geometrical parameters were designed to produce identical degenerate paired photons at 1536\,nm within a bandwidth of 50\,nm when pumped at 768\,nm in the picosecond regime. Since the filter bandwidths are much narrower than the down conversion bandwidth ($\frac{\Delta\lambda_{spdc}}{\Delta\lambda_{filter}}\geq 10^{2}$), experimentally, the two sources are simply independently stabilized at the temperature of 343\,K to within 0.1\,K.

From each source, we select pairs of photons meeting the Fourier transform criterion using narrowband demultiplexers made of two optical circulators and a pair of FBG filters. These are set to reflect energy-matched pairs of wavelength at 1537.4\,nm and 1534.6\,nm, i.e. around degeneracy, associated with bandwidths of 800\,pm and 250\,pm, respectively. A slightly wider (800\,pm) filter is used on the long wavelength (1537\,nm) photons mainly to minimize overall losses. This filtering solution based on standard telecom components provides a clever way to separate deterministically the photons at the output of the waveguides and makes them each available in a single mode fiber~\cite{Halder_Ent_Indep_07}. The 250\,pm filters bandwidth has been choosen for two reasons : (a) it is narrow enough to accommodate the 1.2\,ps-duration of the laser pulses and any possible jitter up to 4\,ps due to laser synchronization and/or dispersion in the fiber; (b) it is large enough to stabilize their central wavelength using two independent proportional-integral-derivative temperature controllers.

\begin{figure}[h]
\scalebox{0.7}{\includegraphics{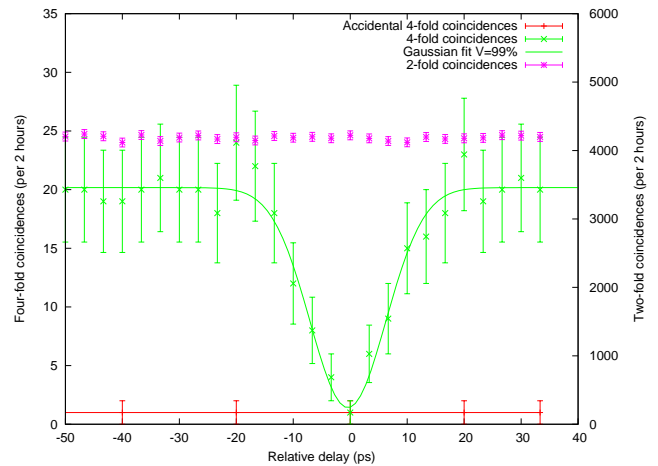}}
\caption{Four-fold coincidence rate as a function of the relative delay, $\delta t$, between the interfering photons ($\sim0.04$ photon/pulse). We clearly observe a dip for $\delta t = 0$ that reaches the noise level. The Gaussian fit of the interference pattern shows a net visibility of $99\%\pm3\%$. We also plotted one of the two-fold coincidences related to one of the sources after the BS to verify this figure is constant.}
\label{Fig_HOM_dip}
\end{figure}

To observe the two-photon interference, four-fold coincidences are detected thanks to four InGaAs-APDs (idQuantique 201) triggered by a Peltier cooled Si-APD (idQuantique 100) placed on the path of an attenuated fraction of the pump beam. Since these InGaAs-APDs are designed to handle a maximum triggering rate of 1\,MHz, this technique allows working at an average clock of 600\,kHz randomly picked from the 76\,MHz laser repetition rate. In our case, faster InGaAs-APD would highly improve the overall coincidence rate as shown in Ref.~\cite{Takesue_Ent_swap_09}. The evolution of the effective four-fold coincidence rate is given in \figurename{\ref{Fig_HOM_dip}} in which a remarkable dip is obtained when the delay between the two inner photons is zero.

\begin{table*}[htbp]
\caption{\label{comparison}Compared two-photon interference visibilities, source brightness, and overall experimental coincidence rates, between reported works in continuous wave (CW), picosecond (ps), and femtosecond (fs) regimes. Regarding time uncertainties, note that the first number is always associated with the pulse duration while the $\pm$ sign corresponds to the synchronisation jitter if two pulsed lasers are involved. In contrast, the time uncertainty for the CW case is associated with the detector's timing jitters. The coherence time is calculated for Gaussian filters using the standard relation $\tau_c=0.44\frac{\lambda^2}{c\Delta\lambda}$.}
\begin{ruledtabular}
\begin{tabular}{c|ccc|ccc|ccc}
Reference     & Source    & Wavelength & Brightness   & Regime\footnotemark[1]     & Filter bandwidth & Rate & Raw & Net \\
& & $nm$ & $pair/s/pm/mW$ & (time uncertainty) & (coherence time) & $pair/s$  & visibility & visibility \\
\hline
Geneva~\cite{Halder_Ent_Indep_07}  & PPLN/w  & 1550  &  $0.9\times10^{3}$ & $2\times$ CW (70\,ps) & 0.01\,nm (350\,ps)  & $3\times10^{-3}$ &  NA & 77\%\\
\textbf{Nice} & \textbf{PPLN/w} & \textbf{1550} & $1.6\times10^{3}$ & \textbf{$1\times$ ps} (\textbf{1.2}\,ps) & \textbf{0.25}\,nm (\textbf{14}\,ps) & $3\times10^{-1}$ & \textbf{93\%} & \textbf{99\%}\\
Atsugi~\cite{Takesue_Ent_swap_09} & Fiber   & 1550  & NA & $1\times$ ps (19\,ps)  & 0.2\,nm (18\,ps)  & $2$\footnotemark[2]  & 64\% & NA\\
Bristol~\cite{Fulconis_TwoPhotInterf_07} & Fiber   & 600  & NA & $1\times$ ps (1.5\,ps)  & 0.3\,nm (1.8\,ps)  & $4\times10^{-1}$  & 88\% & NA\\
Geneva~\cite{DeRiedmatten_Qint_SpatialSeparated_03} & Bulk LBO & 1310 & NA & $1\times$ fs (200\,fs)  & 5\,nm (500\,fs)     & $7\times10^{-1}$ & 77\% & 84\%\\
Beijing~\cite{Yang_Syncro_Indep_06} & Bulk BBO & 800 & $1.2\times10^{-2}$ & $2\times$ fs\footnotemark[3]
(60$\pm$ 2\,fs) & 2.8\,nm (335\,fs) & $3\times10^{-2}$ & 82\% & NA\\
Vienna~\cite{Kaltenbaek_Inter_Indep_09}  & Bulk BBO & 800 &  NA  & $2\times$ fs (50$\pm$ 260\,fs ) & 0.4\,nm (2.3\,ps) &  $1\times10^{-2}$ & 96\% & NA\\
\end{tabular}
\end{ruledtabular}
\begin{minipage}{20cm}
\footnotetext[1]{In this column, the figures associated with the pump regimes correspond to the number of lasers employed.}
\footnotetext[2]{The experiment operates at 500\,MHz repetition rate therefore increasing the available photon pair rate.}
\footnotetext[3]{The two lasers are not truly independent since they share the same Ti:Sapphire crystal Kerr medium for accurate synchronization.}
\end{minipage}
\end{table*}

Using only 1\,mW of mean pump power per source, we obtained $4\times10^{3}$ two-fold coincidences per hour and per source which corresponds to about 0.05 photon per pulse. In terms of normalized brightness it means that our sources reach $1.6\times10^{3}$\,pairs$\cdot$s$^{-1}\cdot$pm$^{-1}\cdot$mW$^{-1}$ after the filtering stage. This brightness corresponds to the state-of-the-art for PPLN waveguides and is basically five orders of magnitude larger than that of bulk crystals~\cite{Tanzilli_PPLNW_02,Halder_HighCoh_pairSource_08}. When the two photons are made indistinguishable in time ($\delta t$) thanks to the adjustment of the retro-reflector (R) placed in front of one sample, a $93\%\pm3\%$ reduction in the raw four-fold coincidence count is obtained. By correctly recording the accidental coincidences, we are able to demonstrate that the net visibility reaches $99\%\pm3\%$. The full width at half maximum of the dip is approximately $\Delta\tau\approx11\,$ps, which is in good agreement with the coherence time of 14\,ps expected from the filter bandwidths. This interference visibility is also extremely close to the maximum value of $99.9\%$ calculated from the theory taking into account our filtering bandwidths, as properly outlined in Refs.~\cite{LegeroRempe_SPcharac_coalecence_03,DeRiedmatten_Qint_SpatialSeparated_03,Rbook}. Note that the very high stability offered by our guided-wave scheme allowed us performing 48h-long measurement leading to more than 30 points evenly spread over the dip.

Table~\ref{comparison} presents the compared results reported in the literature for similar configurations which can be differentiated by four key points: (i) the type of nonlinear generators, (ii) the emitted photon wavelengths, (iii) the pumping regime and its associated time uncertainty, (iv) and, finally, the applied filtering bandwidths and their associated coherence times. From a comparison between the reported results, we conclude that the combination of the picosecond regime and the single mode properties of the employed telecommunication components is the most performant scheme. It allows matching efficiently the inequality~(\ref{temp}) and obtaining near-perfect two-photon interference. For practical long distance quantum communication, it is also interesting to compare the effective number of pairs available per second at Alice and Bob's locations. This figure of merit is calculated by normalizing the four-fold coincidence rates with respect to Alice and Bob's detector efficiencies. A comparison with similar configurations at telecom wavelengths indicates that the picosecond regime allows maintaining a high effective pair production rate equivalent to that of the femtosecond regime but shows a much higher visibility. One should also note the two orders of magnitude rate difference for equivalent source technologies and brightness between the CW and picosecond regimes. In addition, the best visibility reported in Ref.~\cite{Kaltenbaek_Inter_Indep_09} with photons in the visible comes from the fact that the filtering bandwidths are on the order of those necessary for the picosecond regime. Consequently, the overall brightness is very low considering the use of efficient Si-APDs. Finally, bright fiber sources are contaminated by intrinsic spontaneous Raman scattering noise, leading to accidental coincidences and to reduced visibilities~\cite{Takesue_Ent_swap_09,Fulconis_TwoPhotInterf_07}.

The results demonstrated in this experiment are of broad interest since the obtained net visibility is the best value reported to date for comparable configurations. According to reference~\cite{Simon07}, this allows, for the first time, considering the possibility of using quantum relays in actual quantum cryptography networks. In this context, where two synchronized picosecond lasers must be used, we not only expect results of the same order, thanks to the high experimental stability and versatility demonstrated, but also because of the reduced-constraint on the synchronization issue~\cite{Hudson_synchro_06,Kim_synchro_08}.

The authors thank M. P. de Micheli and S. Tascu for the realization of the PPLN wave\-guides, G. Sauder for technical support, and A. Beveratos, H. Zbinden, and K. Thyagarajan for fruitful discussions. This work has been supported by the CNRS, the University of Nice -- Sophia Antipolis, and the French Ministry of Research.

\bibliography{pra_picosecond_dip_resub.bib}

\end{document}